\documentclass[aps,prd,preprint,superscriptaddress,showpacs]{revtex4}
\usepackage{graphicx}

\usepackage{ulem}
\usepackage{color}
\definecolor{My_red}        {cmyk}{0.00,1.00,1.00,0.20}

\newcommand{\bmat}{\left(\begin{array}}
\newcommand{\emat}{\end{array}\right)}
\newcommand{\beq}{\begin{equation}}
\newcommand{\eeq}{\end{equation}}

\def\bwt{\begin{widetext}}
\def\ewt{\end{widetext}}
\def\be{\begin{equation}}
\def\ee{\end{equation}}
\def\bea{\begin{eqnarray}}
\def\eea{\end{eqnarray}}
\def\bean{\begin{eqnarray*}}
\def\eean{\end{eqnarray*}}
\def\bary{\begin{array}}
\def\eary{\end{array}}
\def\bit{\begin{itemize}}
\def\eit{\end{itemize}}

\def\su5u1{SU(5) \times U(1)}
\def\fsu5u1{SU(5) \times U(1)'}
\def\so10{SO(10)}
\def\sq20{SO(10) \times SO(10)}

\def\bwt{\begin{widetext}}
\def\ewt{\end{widetext}}
\def\be{\begin{equation}}
\def\ee{\end{equation}}
\def\bea{\begin{eqnarray}}
\def\eea{\end{eqnarray}}
\def\bean{\begin{eqnarray*}}
\def\eean{\end{eqnarray*}}
\def\bary{\begin{array}}
\def\eary{\end{array}}
\def\bit{\begin{itemize}}
\def\eit{\end{itemize}}

\def\su5u1{SU(5) \times U(1)}
\def\fsu5u1{SU(5) \times U(1)'}
\def\so10{SO(10)}
\def\sq20{SO(10) \times SO(10)}

\usepackage[centertags]{amsmath}
\usepackage{amssymb}

\begin{document}

\title{A Modified Holographic Dark Energy Model with Infrared
 Infinite Extra Dimension(s)}

\author{Yungui Gong}
\email{gongyg@cqupt.edu.cn}
\affiliation{College of Mathematics and Physics, Chongqing
University of Posts and Telecommunications, Chongqing 400065,
China}

\affiliation{Kavli Institute for Theoretical Physics China, Chinese
Academy of Sciences, Beijing 100190, P. R. China}

\author{Tianjun Li}

\affiliation{Key Laboratory of Frontiers in Theoretical Physics,
      Institute of Theoretical Physics, Chinese Academy of Sciences,
Beijing 100190, P. R. China }

\affiliation{George P. and Cynthia W. Mitchell Institute for
Fundamental Physics, Texas A$\&$M University, College Station, TX
77843, USA }

\affiliation{Kavli Institute for Theoretical Physics China, Chinese
Academy of Sciences, Beijing 100190, P. R. China}

\date{12/18/2009}

\begin{abstract}

We propose a modified holographic dark energy (MHDE) model
with the Hubble scale as the infrared (IR) cutoff.
Introducing the infinite extra dimension(s) at very large
distance scale,
we consider the black hole mass in higher dimensions
as the ultraviolet cutoff. Thus, we can probe the
effects of the IR infinite extra dimension(s).
As a concrete example, we consider the
Dvali-Gabadadze-Porrati (DGP) model
 and its generalization. We find that the DGP model
is dual to the MHDE model in five dimensions, and
the $\Lambda$CDM model is dual to the MHDE model
in six dimensions. Fitting the MHDE model to
the observational data, we obtain
that $\Omega_{m0}=0.269^{+0.030}_{-0.027}$, $\Omega_{k0}=0.003^{+0.011}_{-0.012}$, and
the number of the spatial dimensions is
$N=4.78^{+0.68}_{-0.44}$. The best fit value of $N$
implies that there might exist two
IR infinite  extra dimensions.

\end{abstract}

\pacs{98.80.-k,95.36.+x}

\preprint{CAS-KITPC/ITP-125, MIFP-09-28}

\maketitle

\section{Introduction}

 From the cosmological observations such as Type Ia supernova
(SN Ia)~\cite{Riess}, cosmic microwave background (CMB)~\cite{spergel},
 and large scale structure (LSS)~\cite{Tegmark}, there are
strong evidences for  dark energy (DE)
which drives the accelerated expansion of the Universe.
The most naive DE candidate is the cosmological constant (CC)
introduced by Einstein. Although CC is consistent with
the cosmological observations, there exists a fine-tuning
problem~\cite{Weinberg}:
 CC ($(2.3\times 10^{-3} ~{\rm eV})^4$) is extremely small
comparing to the known energy scales such
as the reduced Planck scale $M_{\rm Pl}\equiv 1/\sqrt{8\pi G}$
(about $2.4\times 10^{19}~{\rm GeV}$) in
general relativity and the electroweak scale (about $91~{\rm GeV}$)
in particle physics, and we do not have any symmetry which
can protect the tiny CC against quantum corrections
(supersymmetry must be broken above the electroweak scale).
This fine-tuning problem is the greatest challenge in high energy
physics. Also, there is a coincident problem~\cite{Weinberg}:
why the DE and dark matter energy densities are comparable today
since their evolutions
are different as the Universe expands. Therefore, cosmologists
and particle physicists have proposed some other DE models,
for example, quintessence~\cite{quint}, phantom~\cite{phantom},
$k$-essence~\cite{k}, tachyon~\cite{tachyonic},
quintom~\cite{Feng:2004ad}, hessence~\cite{hessence},
Chaplygin gas~\cite{Chaplygin},
Yang-Mills condensate \cite{YMC}, etc.

The DE problem might be a problem in quantum gravity~\cite{Witten}.
However, we do not have a complete theory of quantum
gravity right now. Fortunately, an important progress in the studies
of the black hole theory and string theory is the proposal of
the holographic principle~\cite{'t Hooft93}, which
 may be considered as a fundamental principle
of quantum gravity and then shed some light on the DE problem.
Using the effective quantum field theory, Cohen, Kaplan and
Nelson suggested that the quantum zero-point energy
of a system with size $L$ should not exceed the mass of a black hole
with the same size, {\it i.e.}, $L^3\rho_{\rm V}\leq LM_{\rm Pl}^2$, where
$\rho_{\rm V}$ is the quantum zero-point energy density~\cite{Cohen:1998zx}.
Thus, the ultraviolet (UV) cutoff scale of a system is connected to
its infrared (IR) cutoff scale. Applying this idea to the whole
Universe, we can consider the vacuum energy as DE
with density $\rho_{\rm DE} \equiv \rho_{\rm V}$.
Choosing the largest IR cutoff $L$ which saturates the
inequality,  we obtain the holographic DE density
\begin{equation}
\rho_{DE}=3c^2 M_{\rm Pl}^2L^{-2}~,
\label{DE}
\end{equation}
where $c$ is an unknown constant due to the theoretical uncertainties
and can only be determined by observations. Interestingly,
taking $L$ as the size of the current Universe which is
 the Hubble radius $H^{-1}$, one finds that the DE
density is close to the observed value. However,
Hsu~\cite{Hsu} pointed out that this yields a wrong equation
of state for DE.

To solve this problem, Li~\cite{Li} chose the future event
horizon of the Universe as the IR cutoff, and found that the model
is a viable DE model. However, there exists
an obvious draw back concerning causality: the event horizon is
a global concept of space-time, and the existence of
event horizon depends on the future evolution of the Universe,
{\it i.e.}, the event horizon exists if and only if the
Universe is accelerating. So, the original holographic DE model
has presumed the existence of accelerated expansion. To avoid the causality problem,
Cai proposed  the agegraphic DE model
in which the age of the Universe can
be chosen as the IR cutoff~\cite{Cai}.
 A new version of this model, which
replaces the age of the Universe by the conformal age of the
Universe, was suggested as well~\cite{Wei}.
Moreover, Gao, Wu, Chen, and Shen proposed the
holographic Ricci DE model where
the  average radius of the Ricci
scalar curvature is chosen as the IR cutoff~\cite{Gao}.
The phenomenological consequences of these
models have been studied extensively~\cite{refHDE,refADE,refRDE,Li:2009bn}.

In this Letter, to obtain the accelerating Universe and
solve the causality problem, we propose a modified
holographic DE (MHDE) model with the Hubble scale as
the IR cutoff, and the UV and IR connection
is modified by using the black hole mass in higher dimensions.
As a concrete example, we consider the Dvali-Gabadadze-Porrati (DGP) models
which have infinite extra dimension(s)~\cite{dgp, dg}.
The characteristic distance scale in such theories is
called the crossover scale $r_c$ (for definitions,
please see Section II.). At the distances
that are smaller than $r_c$, we observe  four-dimensional gravity.
While at the distances larger than $r_c$, we observe
higher-dimensional gravity. So at the IR energy scale
which is smaller than $r_c^{-1}$ (or say at the distance
larger than $r_c$), we need to consider
the effect of extra dimensions. In particular, the mass of
the higher-dimensional black hole is different from that
of the four-dimensional one. Therefore,
the UV cutoff scale in the MHDE model is different from
that in the original holographic DE model.
Moreover, our model provides a way to probe
the IR infinite extra dimensions:
the DGP model is dual to the MHDE model in five dimensions,
and  the $\Lambda$CDM model is dual to the MHDE model
in six dimensions.
Interestingly, with the Hubble scale
as the IR cutoff scale, we can not only obtain the observed
DE density with correct equation of state, but also
 avoid the causality problem.
The best fit value
indicates that there might exist two
IR infinite  extra dimensions.

This Letter is organized as follows.  We present our model
in Section II. And we discuss the observational constraints in
Section III. Our conclusion is in Section IV.

%%%%%%%%%%%%%%%%%%%%%%%%%%%%%%%%%%%%%%%%%%%%%%%%%%%%%%%%%%%%%%%%%%%%%%%%%%%%%%%%

%%%%%%%%%%%%%%%%%%%%%%%%%%%%%%%%%%%%%%%%%%%%%%%%%%%%%%%%%%%%%%%%%%%%%%%%%%%%%%%%

\section{The model}

First, let us briefly review the DGP model and its
generalization~\cite{dgp, dg}.
The theory is a brane-world model which is embedded in a space-time
with (asymptotically) flat infinite  $n$ extra  space dimensions.
All the standard model (SM) particles are localized on the D3-branes, and the
corresponding cut-off scale on the observable D3-branes
can be the grand unification scale
or higher. Also, the gravity spreads over the whole $4+n$ space-time
dimensions. Thus, the action is
\begin{equation}
S ~=~ {{M_*^{2+n}}\over 2}
\int d^4x d^ny {\sqrt G} {\cal R} + \int d^4x
{\sqrt g} \left(T_0 + {{M^2_{\rm Pl}}\over 2} R
+ {\cal L}_{\rm SM} \right)~,~\,
\end{equation}
where $x^{\mu}$ and $y^i$ are respectively the coordinates for the
Minkowski space-time and the extra space dimensions,
$G=|{\rm det}(G_{AB})|$ and $G_{AB}$ is the metric for
the whole space-time, $g=|{\rm det}(g_{\mu\nu})|$
and $g_{\mu\nu}$ is the reduced metric on the D3-branes,
 $M_*$ is the
high-dimensional Planck scale, $T_0$ is the brane tension, and
${\cal L}_{\rm SM}$ is the SM Lagrangian~\cite{dgp, dg}.
The characteristic distance scale in such theories is
called the crossover scale. For five-dimensional
space-time with $n=1$, the crossover scale is~\cite{dgp}
\begin{equation}
r_c ~\sim~ {{M_{\rm Pl}^2}\over {M_*^3}}~.~\,
\end{equation}
And for six- or higher-dimensional space-time with zero tension
branes, {\it i.e.}, $n\ge 2$, the crossover scale is~\cite{Dvali:2001ae}
\begin{equation}
r_c ~\sim~ {{M_{\rm Pl}}\over {M_*^2}}~.~\,
\end{equation}
Interestingly, at distance smaller than $r_c$,
{\it i.e.}, $r < r_c$, we observe the
four-dimensional gravity, while at distance $r > r_c$, we
indeed observe high-dimensional gravity. Moreover,
from the data on
sub-millimeter gravity measurements~\cite{adel} and
the accelerator, astrophysical and cosmological
data~\cite{Dvali:2001gm},
the lower bound on the value of $M_*$ is $M_* \ge 10^{-3}~{\rm eV}$.
Also, from the solar system measurements of Newton's law,
one obtains the upper bound on $M_*$: $M_* \le 10^{12}~{\rm eV}$
for five-dimensional space-time, and $M_* \le 10^{4}~{\rm eV}$
for six- and higher-dimensional
space-time~\cite{Dvali:2001ae, Dick:2001np}.

The idea of holography may be used to
solve the CC problem \cite{Cohen:1998zx,lvdark}.
Cohen, Kaplan and Nelson proposed that for any state with energy $E$ in the
Hilbert space, the corresponding Schwarzschild
radius $R_s\sim E$ is less than the IR cutoff $L$
\cite{Cohen:1998zx}. Under this assumption, a relationship between the
 UV cutoff $\rho_D^{1/4}$ and the IR cutoff is derived,
{\it i.e.}, $L^3\rho_d\sim L$ \cite{Cohen:1998zx}.
Here the relationship $M\sim R_s$ between the mass and the horizon of
the Schwarzschild black hole in four dimensions is used. In $N+1$ dimensions
with $N\equiv n+3$,
the mass of the Schwarzschild black hole is \cite{myers}
\begin{equation}
\label{nbkhole}
M=\frac{(N-1)\Omega_{N-1}}{16\pi G_N}r^{N-2}_H~,~\,
\end{equation}
where $8\pi G_N \equiv M_*^{1-N}$.
If we can see the effect of extra dimensions, then we have the relation
\begin{equation}
\label{uvir}
L^3\rho_d\sim \frac{(N-1)\Omega_{N-1}}{16\pi G_N}L^{N-2},
\end{equation}
and the DE density
\begin{equation}
\label{mholrd}
\rho_d= \frac{d(N-1)\Omega_{N-1}}{16\pi G_N}L^{N-5},
\end{equation}
where $d$ is an unknown constant accounting for theoretical uncertainties.
If we choose the Hubble horizon as the IR cutoff,
then we get the MHDE density
\begin{equation}
\label{mholrhod}
\rho_d= \frac{d(N-1)\Omega_{N-1}}{16\pi G_N}H^{5-N}.
\end{equation}
The  mass of the Schwarzschild black hole is
\begin{equation}
\label{nbkhole}
M=\frac{(N-1)\Omega_{N-1}}{2} M_*^{N-1}r_H^{N-2}~.~\,
\end{equation}
Thus, the high-dimensional black hole mass can be much higher
than the Planck scale in our scenarios. However, to compare
with particle physics theory, we consider the Lagrangian density.
In our models, we have the dark energy density as follows
\begin{equation}
\label{mholrhod}
\rho_d= \frac{d(N-1)\Omega_{N-1}}{2} M_*^{N-1}L^{N-5}.
\end{equation}
Thus, for the model with one or two extra dimensions, the
dark energy density
is proportional to $M_*^3 L^{-1}$ and $M_*^4$, respectively.
Thus, they are consistent with particle physics theory. However,
for the models with three or more extra dimensions, the
dark energy density is proportional to $ M_*^{N-1}L^{N-5}$
with $N \ge 6$, thus, these models may have pretty
large dark energy density, which is not consistent with
particle physics. Therefore, we only consider one or
two extra dimensions, {\it i.e.}, the space-time dimensions
are five and six, respectively.
In particular, for five-dimensional  and six-dimensional space-time,
we obtain the observed DE density for
$M_* \sim 10^{7}~{\rm eV}$ and $M_* \sim 10^{-3}~{\rm eV}$, respectively.
%%% Note that $M_* \ge 10^{-3}~{\rm eV}$, to explain the
%%% observed dark energy density,
%%% we can only consider one and two infinite large extra dimensions,
%%% {\it i.e.}, $n=$1 and 2, or $N=4$ and $5$.
Now the Friedmann equation becomes
\begin{equation}
\label{freq1}
H^2+\frac{k}{a^2}=\frac{8\pi G}{3} (\rho_m+\rho_r+\rho_d),
\end{equation}
where the matter energy density $\rho_m=\rho_{m0}(a_0/a)^3$
and the radiation energy density $\rho_r=\rho_{r0}(a_0/a)^4$.
Substituting Eq. (\ref{mholrhod}) into Eq. (\ref{freq1}), we get
\begin{equation}
\label{freq2}
\left(\frac{H}{H_0}\right)^2-(1-\Omega_{m0}-\Omega_{k0}-\Omega_{r0})\left(\frac{H}{H_0}\right)^{5-N}=
\Omega_{m0}(1+z)^3+\Omega_{r0}(1+z)^4+\Omega_{k0}(1+z)^2.
\end{equation}
This is the same as the $\alpha$ dark energy model proposed
 by Dvali and Turner in Ref.~\cite{turner} where $\alpha=5-N$.
The crossover scale in this model is
 \begin{equation}
 \label{mholrc}
 r_c\sim \frac{M_{\rm Pl}^{2/(N-3)}}{M_*^{(N-1)/(N-3)}}~.~\,
 \end{equation}
In~\cite{turner}, the model is proposed as
the modification to the Friedmann equation motivated
by extra dimensions, and it was found that
$\alpha\lesssim 1$ to be consistent with astronomical observations.
Here we use the holographic idea to derive the
above Friedmann equation and take the model as a dark energy model.
We also connect the
holographic idea with the effect of IR infinite extra dimensions.
Furthermore, the parameter $\alpha$ has the physical meaning of the
number of the spatial dimensions.
If $N=4$, {\it i.e.} , with one extra dimension, we
recover the DGP model for the flat case. So the DGP model may
be interpreted as the holographic dark energy model with the Hubble
scale as the IR cutoff.  When $N=5$, {\it i.e.}, with
two extra dimensions, we recover the standard $\Lambda$CDM model.
In other words, the cosmological constant is a special case of our model, and
it may be interpreted as the effect
of two extra dimensions. In our model, we choose the Hubble scale as
the IR cutoff, so the problem of circular reasoning is avoided. In addition, the
choice of the Hubble scale as the IR cutoff is more natural than other choices.

Combining the Friedmann equation (\ref{freq1}) with the energy conservation,
we get the equation of state parameter $w_d$
for the MHDE
\begin{equation}
\label{wdeq}
w_d=-1+\frac{2\alpha(1-\Omega_d)}{3(2-\alpha\Omega_d)}+\frac{\alpha(\Omega_m+2\Omega_r)}{3(2-\alpha\Omega_d)}.
\end{equation}
So $w_d$ changes from $-1+2\alpha/3$ during the radiation dominated era to $-1+\alpha/2$ during
the matter dominated era. The deceleration parameter $q=-\ddot{a}/(aH^2)$ is
\begin{equation}
\label{qeq}
q=-1+\frac{2\Omega_k+3\Omega_m+4\Omega_r}{2-\alpha\Omega_d}.
\end{equation}
Using the Friedmann equation $\Omega_m+\Omega_r+\Omega_k+\Omega_d=1$ and $q=0$,
we obtain at the transition redshift $z_t$
\begin{equation}
\label{zteq1}
(N-3)\Omega_k(z_t)+(N-2)\Omega_m(z_t)+(N-1)\Omega_r(z_t)=N-3.
\end{equation}
Since the transition from deceleration to  acceleration happened very recently,
 we can ignore the contribution due to the radiation, and then we have
\begin{equation}
\label{zteq}
(N-3)\Omega_{k0}(1+z_t)^2+(N-2)\Omega_{m0}(1+z_t)^3=(N-3)(H/H_0)^2.
\end{equation}

\section{Observational constraints}

Now we use the observational data to fit the MHDE model. The parameters in the model are
determined by minimizing
$\chi^2=\chi^2_{sn}+\chi^2_{bao}+\chi^2_{cmb}+\chi^2_h$. For the
SN Ia data, we use the Constitution compilation of 397
SN Ia \cite{consta}. The Constitution sample adds 185 CfA3 SN Ia data to the Union sample \cite{union}.
The addition of CfA3 sample increases the number of nearby SN Ia by a factor
of roughly $2.6-2.9$ and reduces the statistical uncertainties.
The Union compilation has 57 nearby SN Ia
and 250 high-$z$ SN Ia. It includes the Supernova Legacy
Survey \cite{astier} and the ESSENCE Survey \cite{riess,essence},
the older observed SN Ia data, and the extended data set of distant SN Ia
observed with the Hubble space telescope. To fit the SN Ia
data, we define
\begin{equation}
\label{chi}
\chi^2_{sn}=\sum_{i=1}^{307}\frac{[\mu_{obs}(z_i)-\mu(z_i)]^2}{\sigma^2_i},
\end{equation}
where the extinction-corrected distance modulus
$\mu(z)=5\log_{10}[d_L(z)/{\rm Mpc}]+25$, $\mu_{obs}$ is the observed distance modulus,
 $\sigma_i$ is the total
uncertainty in the SN Ia data, and the luminosity distance is
\begin{equation}
\label{lum}
d_L(z)=\frac{1+z}{H_0\sqrt{|\Omega_{k}|}} {\rm
sinn}\left[\sqrt{|\Omega_{k}|}\int_0^z
\frac{dz'}{E(z')}\right],
\end{equation}
where
\begin{equation}
\frac{{\rm sinn}(\sqrt{|\Omega_k|}x)}{\sqrt{|\Omega_k|}}=\begin{cases}
\sin(\sqrt{|\Omega_k|}x)/\sqrt{|\Omega_k|},& {\rm if}\ \Omega_k<0,\\
x, & {\rm if}\  \Omega_k=0, \\
\sinh(\sqrt{|\Omega_k|}x)/\sqrt{|\Omega_k|}, & {\rm if}\  \Omega_k>0,
\end{cases}
\end{equation}
and the dimensionless Hubble parameter
$E(z)=H(z)/H_0$. In particular, $E(z)=[\Omega_{m0}(1+z)^3+1-\Omega_{m0}]^{1/2}$
for the flat $\Lambda$CDM model and
$E(z)=[\Omega_{m0}(1+z)^3+(1-\Omega_{m0})^2/4]^{1/2}+(1-\Omega_{m0})/2$ for
the flat DGP model. We marginalized over the nuisance parameter $H_0$ when evaluating
$\chi^2_{sn}$.

To use the baryon acoustic oscillation (BAO) measurement from the Sloan digital sky survey data, we define
\cite{sdss6}
\begin{equation}
\label{baochi2}
\chi^2_{bao}=\left(\frac{r_s(z_d)/D_V(z=0.2)-0.198}{0.0058}\right)^2
+\left(\frac{r_s(z_d)/D_V(z=0.35)-0.1094}{0.0033}\right)^2,
\end{equation}
where the effective distance is
\begin{equation}
\label{dvdef}
D_V(z)=\left[\frac{d_L^2(z)}{(1+z)^2}\frac{z}{H(z)}\right]^{1/3}.
\end{equation}
The redshift $z_d$ is fitted with the formulas \cite{hu98}
\begin{equation}
\label{zdfiteq}
z_d=\frac{1291(\Omega_{m0} h^2)^{0.251}}{1+0.659(\Omega_{m0} h^2)^{0.828}}[1+b_1(\Omega_b h^2)^{b_2}],
\end{equation}
\begin{equation}
\label{b1eq}
b_1=0.313(\Omega_{m0} h^2)^{-0.419}[1+0.607(\Omega_{m0} h^2)^{0.674}],\quad b_2=0.238(\Omega_{m0} h^2)^{0.223},
\end{equation}
and the comoving sound horizon is
\begin{equation}
\label{rshordef}
r_s(z)=\int_z^\infty \frac{dz'}{c_s(z')E(z')},
\end{equation}
where the sound speed $c_s(z)=1/\sqrt{3[1+\bar{R_b}/(1+z)}]$,
the dimensionless Hubble constant $h=H_0/100$,
and $\bar{R_b}=315000\Omega_b h^2(T_{cmb}/2.7{\rm K})^{-4}$.

To implement the Wilkinson microwave anisotropy probe 5 year (WMAP5) data, we need to add three fitting parameters
$R$, $l_a$ and $z_*$, so $\chi^2_{cmb}=\Delta x_i {\rm
Cov}^{-1}(x_i,x_j)\Delta x_j$, where $x_i=(R,\ l_a,\ z_*)$ denotes
the three parameters for WMAP5 data, $\Delta x_i=x_i-x_i^{obs}$ and
Cov$(x_i,x_j)$ is the covariance matrix for the three parameters
\cite{wmap5}. The acoustic scale $l_A$ is
\begin{equation}
\label{ladefeq}
l_A=\frac{\pi d_L(z_*)}{(1+z_*)r_s(z_*)},
\end{equation}
where the redshift $z_*$ is given by \cite{hu96}
\begin{equation}
\label{zstareq}
z_*=1048[1+0.00124(\Omega_b h^2)^{-0.738}][1+g_1(\Omega_{m0} h^2)^{g_2}]=1090.04\pm 0.93,
\end{equation}
\begin{equation}
g_1=\frac{0.0783(\Omega_b h^2)^{-0.238}}{1+39.5(\Omega_b h^2)^{0.763}},\quad
g_2=\frac{0.560}{1+21.1(\Omega_b h^2)^{1.81}}.
\end{equation}
The shift parameter is
\begin{equation}
\label{shift}
R=\frac{\sqrt{\Omega_{m0}}}{\sqrt{|\Omega_{k}|}}{\rm
sinn}\left(\sqrt{|\Omega_{k}|}\int_0^{z_*}\frac{dz}{E(z)}\right)=1.710\pm 0.019.
\end{equation}

Simon, Verde, and Jimenez obtained the Hubble parameter $H(z)$ at
nine different redshifts from the differential ages of passively
evolving galaxies \cite{hz1}. Recently, the authors in \cite{hz2}
obtained $H(z=0.24)=79.69\pm 2.32$, $H(z=0.34)=83.8\pm 2.96$, and $H(z=0.43)=86.45\pm 3.27$ by
taking the BAO scale as a standard ruler in the radial direction. To
use these 12 $H(z)$ data, we define
\begin{equation}
\label{hzchi}
\chi^2_h=\sum_{i=1}^{12}\frac{[H_{obs}(z_i)-H(z_i)]^2}{\sigma_{hi}^2},
\end{equation}
where $\sigma_{hi}$ is the $1\sigma$ uncertainty in the $H(z)$ data.
We also add the prior $H_0=74.2\pm 3.6$ km/s/Mpc given by Riess {\it
et al.} \cite{riess09}. The likelihood for the parameters in the
model and the nuisance parameters $\Omega_b h^2$ and $H_0$ ($h$) is
computed using a Monte Carlo Markov chain (MCMC). The MCMC method
randomly chooses values for the above parameters, evaluates $\chi^2$
and determines whether to accept or reject the set of parameters
using the Metropolis-Hastings algorithm. The set of parameters that
is accepted to the chain forms a new starting point for the next
process, and the process is repeated for a sufficient number of
steps until the required convergence is reached. Our MCMC code is
based on the publicly available package COSMOMC \cite{cosmomc}.

\begin{figure}[htp]
\centering
\includegraphics[width=8cm]{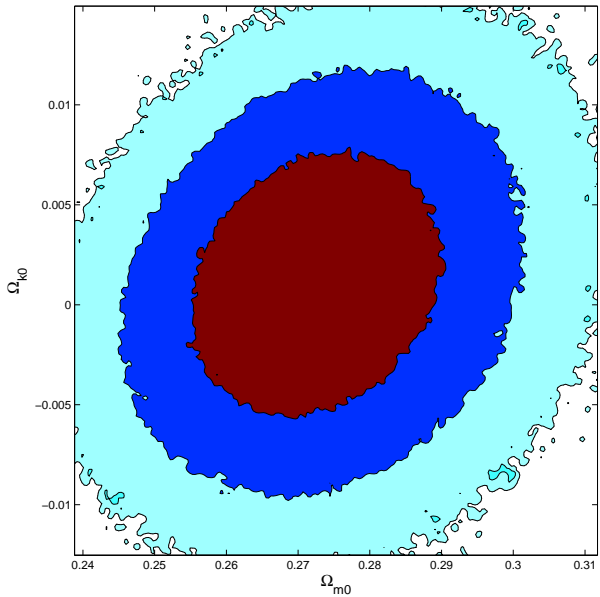}
\caption{The marginalized contours of $\Omega_{m0}$ and $\Omega_{k0}$ in the curved $\Lambda$CDM model.}
\label{omokcont}
\end{figure}

\begin{figure}[htp]
\centering
\includegraphics[width=8cm]{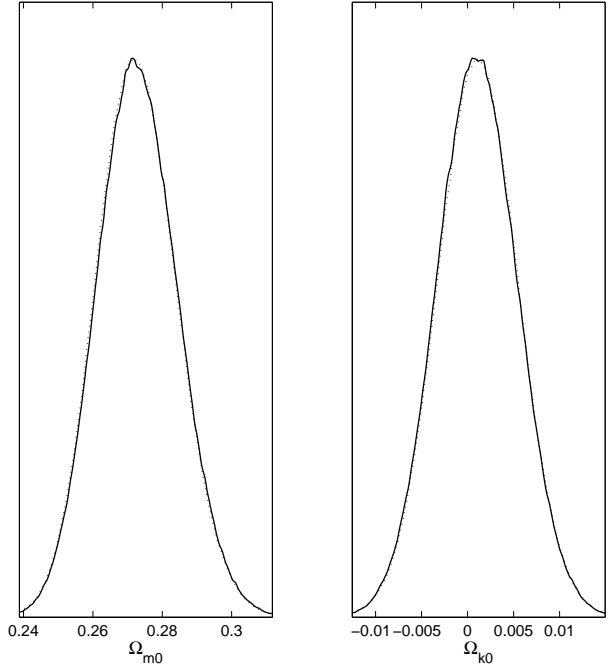}
\caption{The probability distributions of $\Omega_{m0}$ and $\Omega_{k0}$ in the curved $\Lambda$CDM model.}
\label{obsclcdm}
\end{figure}

\begin{figure}[htp]
\centering
\includegraphics[width=8cm]{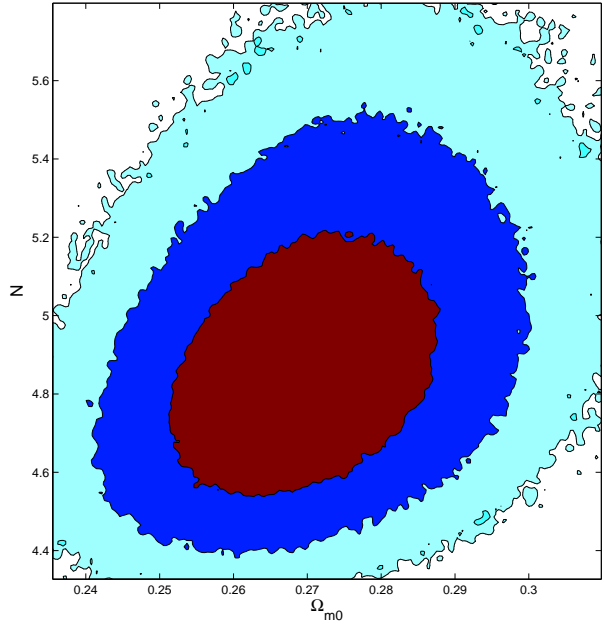}
\caption{The marginalized contours of $\Omega_{m0}$ and $N$ in the flat MHDE model.}
\label{omNcont}
\end{figure}

\begin{figure}[htp]
\centering
\includegraphics[width=8cm]{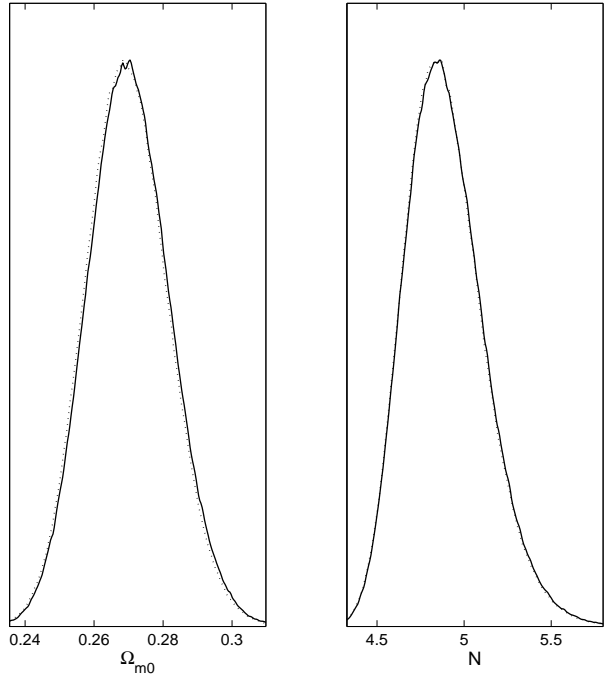}
\caption{The probability distributions of $\Omega_{m0}$ and $N$ in the flat MHDE model.}
\label{obsmhol}
\end{figure}

\begin{figure}[htp]
\centering
\includegraphics[width=8cm]{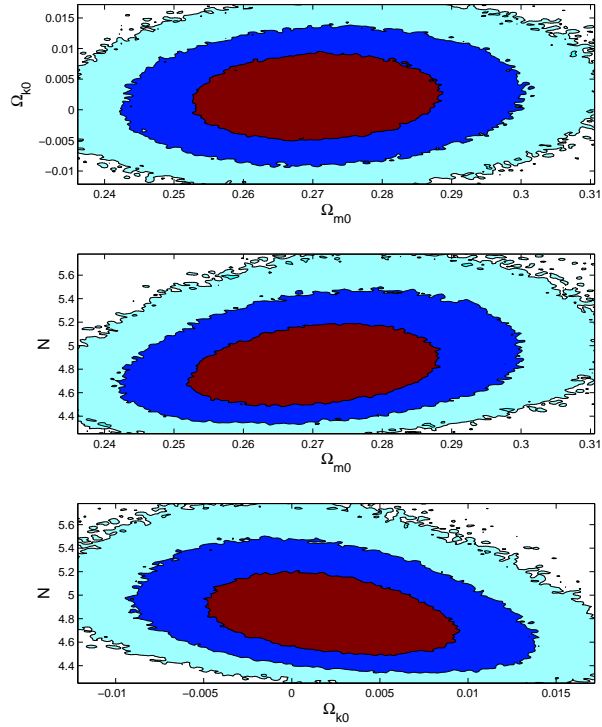}
\caption{The marginalized contours of the parameters in the curved MHDE model.}
\label{cmholcont}
\end{figure}

\begin{figure}[htp]
\centering
\includegraphics[width=8cm]{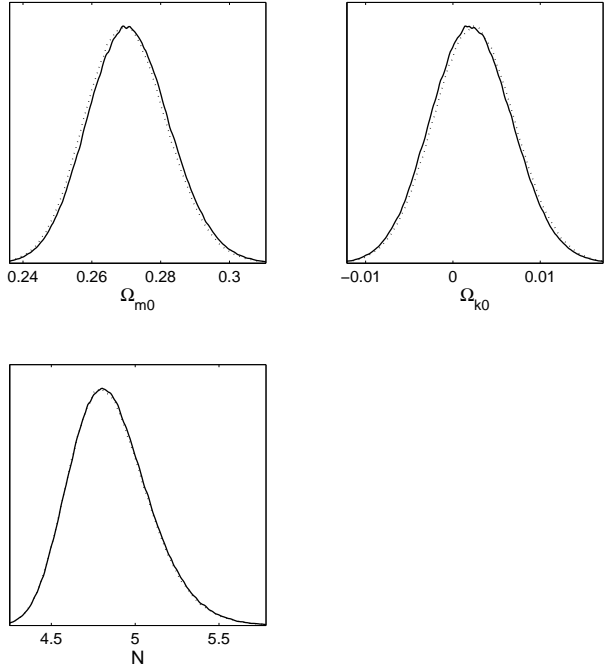}
\caption{The probability distributions of the parameters in the curved MHDE model.}
\label{obscmhol}
\end{figure}

\begin{figure}[htp]
\centering
\includegraphics[width=8cm]{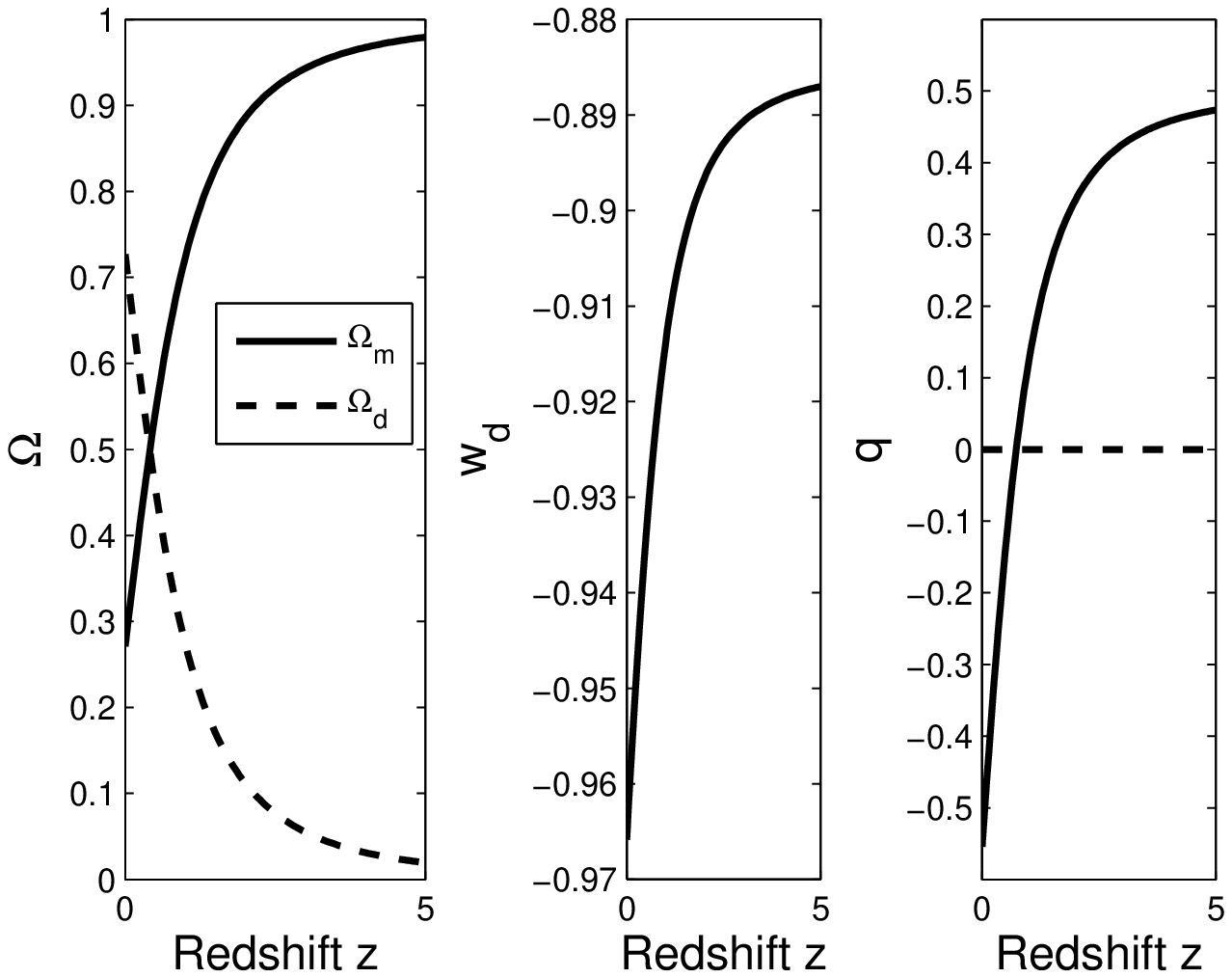}
\caption{The left panel shows the evolutions
of the matter energy density $\Omega_m$ and the holographic DE density $\Omega_d$,
the middle panel shows the evolution of the holographic DE equation of state parameter $w_d$,
and the right panel shows the evolution of the deceleration parameter $q$.}
\label{mholcosm}
\end{figure}

By fitting the flat $\Lambda$CDM model to the combined SN Ia, BAO, WMAP5 and $H(z)$ data, we find
that $\chi^2=483.0$ and $\Omega_{m0}=0.272\pm 0.021$.
If we fit the observational data to the curved $\Lambda$CDM model, we find that $\chi^2=482.9$, $\Omega_{m0}=0.272^{+0.026}_{-0.024}$
and $\Omega_{k0}=0.001_{-0.009}^{+0.010}$. The contour plots and the probability distributions are shown in
Figs. \ref{omokcont} and \ref{obsclcdm}. By fitting the observational data to the flat MHDE model, we find that
$\chi^2=482.5$, $\Omega_{m0}=0.269^{+0.027}_{-0.025}$ and $N=4.84^{+0.56}_{-0.40}$. The contour
plots and the probability distributions are shown in
Figs. \ref{omNcont} and \ref{obsmhol}. Substituting the best fit values $\Omega_{m0}$ and $N$ to Eq. (\ref{zteq}),
we get the transition redshift $z_t=0.746$.
By fitting the observational data to the curved MHDE model, we find that
$\chi^2=482.2$, $\Omega_{m0}=0.269^{+0.030}_{-0.027}$, $\Omega_{k0}=0.003^{+0.011}_{-0.012}$ and $N=4.78^{+0.68}_{-0.44}$.
The contour plot and the probability distributions are shown in
Figs. \ref{cmholcont} and \ref{obscmhol}.  By using the best fit values of $\Omega_{m0}$,
$\Omega_{k0}$, and $N$, we find that the age of the Universe is $t_0=13.74$ Gyr which is consistent
with the result given by WMAP5 \cite{wmap5}.
Substituting the best fit values of $\Omega_{m0}$,
$\Omega_{k0}$, and $N$ to Eq. (\ref{zteq}), we get the transition redshift $z_t=0.739$.
Since the best fit value of
$\Omega_{k0}$ is very small, the spatial geometry of the Universe
is almost flat, so the transition redshift $z_t$
is almost the same for the curved and flat cases.
In other words, we can neglect the contribution due to the curvature term.
By using Eqs. (\ref{wdeq}) and (\ref{qeq}) with the best fit parameter values,
we plot the evolutions of the $\Omega_m$ and $\Omega_d$, the equation of state parameter $w_d$,
and the deceleration parameter $q$ in Fig. \ref{mholcosm}.

%%%%%%%%%%%%%%%%%%%%%%%%%%%%%%%%%%%%%%%%%%%%%%%%%%%%%%%%%%%%%%%%%%%%%%%%%%%%%%%%

%%%%%%%%%%%%%%%%%%%%%%%%%%%%%%%%%%%%%%%%%%%%%%%%%%%%%%%%%%%%%%%%%%%%%%%%%%%%%%%%

\section{Conclusions}

We proposed the MHDE model with the Hubble scale as the IR cutoff,
and our model
fits the observational data as well as that of the $\Lambda$CDM model.
The black hole mass
in higher dimensions is used to modify the UV and IR relation, and
then to derive our model.
Since we used the Hubble scale as the IR cutoff, our model
avoids the problem of circular reasoning.
Furthermore, our model suggests a way probing IR infinite
extra dimensions since
the DGP model and the $\Lambda$CDM model are special cases of our model.
In particular, the DGP model
is dual to the MHDE model in five dimensions, and
the $\Lambda$CDM model is dual to the MHDE
in six dimensions. In other words, the effect of one extra
dimension is manifested as the DGP model, {\it i.e.},
the effect of five-dimensional gravity at large distance can
be seen as DE in four dimensions. Also, the effect
of six-dimensional gravity at large distance can be
considered as the CC
in four dimensions. Fitting the model
to the combined SN Ia, BAO, WMAP5 and $H(z)$ data, we find
that $\Omega_{m0}=0.269^{+0.030}_{-0.027}$, $\Omega_{k0}=0.003^{+0.011}_{-0.012}$ and $N=4.78^{+0.68}_{-0.44}$.
By using the best fit values of $\Omega_{m0}$,
$\Omega_{k0}$, and $N$, we find that the age of the Universe is $t_0=13.74$ Gyr.
The best fit value of $N$ suggests
that there may exist two IR infinite extra dimensions.

%The DGP model is the
%realization of the model in five dimensions.
%How to realize the model in six and higher dimensions is an open question.

\begin{acknowledgments}

We wish to acknowledge the hospitality of the KITPC under their program
``Connecting Fundamental Theory with Cosmological Observations'' during
which this project was initiated.
Y.G. was partially supported by the 973 Program under grant No. 2010CB833004,
by the National Natural Science Foundation of China key project under grant No. 10935013, and by the Chongqing
Science and Technology Commission under grant No. 2009BA4050.
T.L. was supported in part
by the Cambridge-Mitchell Collaboration in Theoretical Cosmology,
by the National Natural Science Foundation
of China  under grant No. 10821504, and by the Chinese
Academy of Sciences under the grant No. KJCX3-SYW-N2.
In addition, this work is supported in part by
the Project of Knowledge Innovation Program (PKIP) of the Chinese
Academy of Sciences under the grant No. KJCX2.YW.W10.

\end{acknowledgments}

%%%%%%%%%%%%%%%%%%%%%%%%%%%%%%%%%%%%%%%%%%%%%%%%%%%%%%%%%%%%%%%%%%%%%%%%%%%%%%%%

%%%%%%%%%%%%%%%%%%%%%%%%%%%%%%%%%%%%%%%%%%%%%%%%%%%%%%%%%%%%%%%%%%%%%%%%%%%%%%%%

%%%%%%%%%%%%%%%%%%%%%%%%%%%%%%%%%%%%%%%%%%%%%%%%%%%%%%%%%%%%%%%%%%%%%%%%%%%%%%%%

%%%%%%%%%%%%%%%%%%%%%%%%%%%%%%%%%%%%%%%%%%%%%%%%%%%%%%%%%%%%%%%%%%%%%%%%%%%%%%%%

\end{document}